\documentclass[preprint]{vgtc}            

\onlineid{1001}

\vgtccategory{Short Paper}

\title{When Provenance Aids and Complicates Reproducibility Judgments}

\author{%
  \authororcid{David Koop}{0000-0002-1825-0097}
}

\authorfooter{
  \item
  	David Koop is with Northern Illinois University.
  	E-mail: dakoop@niu.edu
}

\abstract{
It is well-established that the provenance of a scientific result is important, sometimes more important than the actual result. For computational analyses that involve visualization, this provenance information may contain the steps involved in generating visualizations from raw data. Specifically, data provenance tracks the lineage of data and process provenance tracks the steps executed. In this paper, we argue that the utility of computational provenance may not be as clear-cut as we might like. One common use case for provenance is that the information can be used to reproduce the original result. However, in visualization, the goal is often to communicate results to a user or viewer, and thus the insights obtained are ultimately most important. Viewers can miss important changes or react to unimportant ones. Here, interaction provenance, which tracks a user's actions with a visualization, or insight provenance, which tracks the decision-making process, can help capture what happened but don't remove the issues. In this paper, we present scenarios where provenance impacts reproducibility in different ways. We also explore how provenance and visualizations can be better related.
}


\teaser{
  \centering
  \subfloat[Published Visualization]{
    \includegraphics[width=0.24\textwidth]{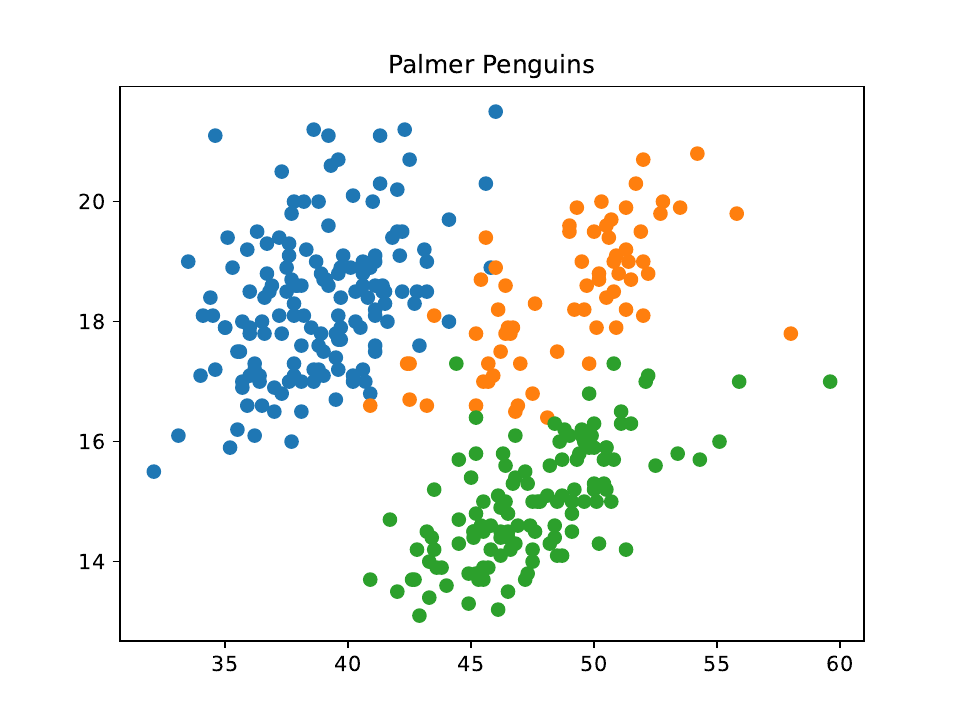}
    \label{fig:published}
  }
  \subfloat[Reproduced Visualization]{
    \includegraphics[width=0.24\textwidth]{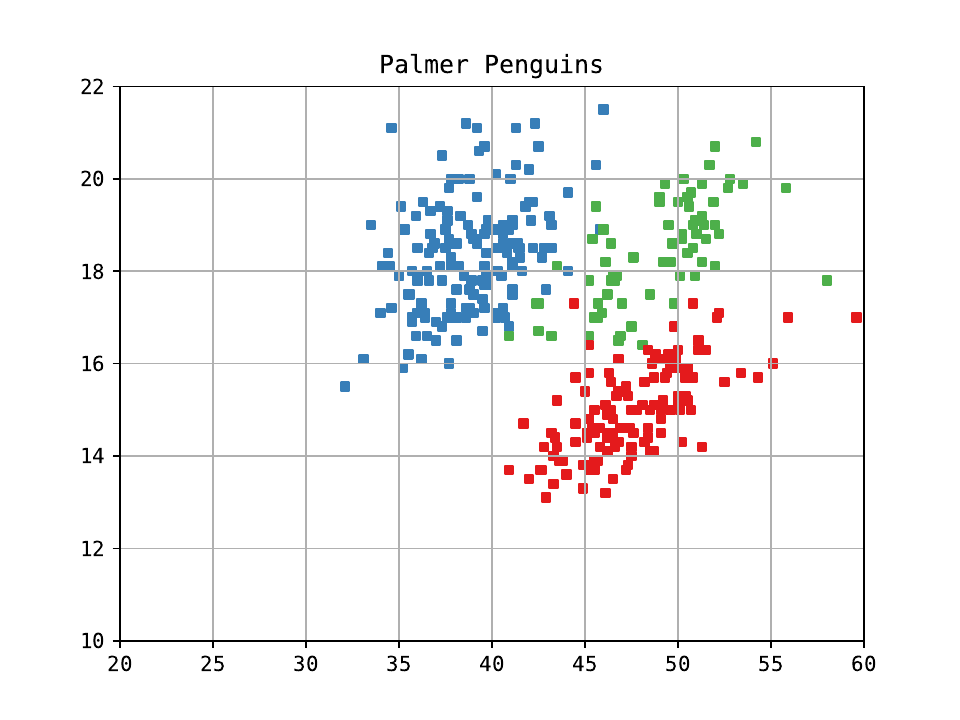}
    \label{fig:reproduced} 
  }
  \subfloat[Modified Data]{
    \includegraphics[width=0.24\textwidth]{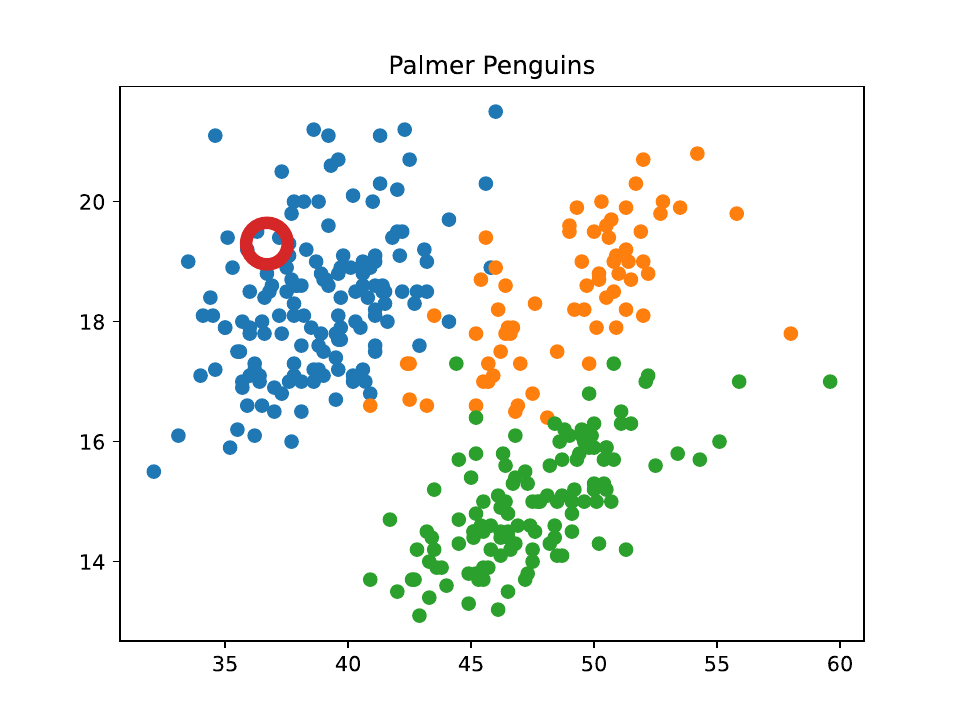}
    \label{fig:modified-data}
  }
  \subfloat[Modified Process]{
    \includegraphics[width=0.24\textwidth]{figures/orig.pdf}
    \label{fig:modified-process}
  }
  \caption{When reproducing a visualization \protect\subref{fig:published}, if default settings are not captured in the provenance, an execution of the same code may generate a very different visualization \protect\subref{fig:reproduced}. When data is modified, it may be difficult to spot the missing point \protect\subref{fig:modified-data}. Finally, a modified process (e.g. different code) can produce the exact same visualization \protect\subref{fig:modified-process}. Provenance may miss changes in~\protect\subref{fig:reproduced}, help detect the changes in~\protect\subref{fig:modified-data}, and be disregarded in~\protect\subref{fig:modified-process}.}
  \label{fig:teaser}
}

\graphicspath{{figs/}{figures/}{pictures/}{images/}{./}} 

\usepackage{tabu}                      
\usepackage{booktabs}                  
\usepackage{lipsum}                    
\usepackage{mwe}                       

\usepackage{mathptmx}                  

\begin{document}


\firstsection{Introduction}

\maketitle

In science, the reproducibility of published work is important. There have been many important studies and high-profile perspectives written about the reproducibility crisis where important results cannot be reproduced later on~\cite{baker:repro,peng:crisis,hutson:ai-repro-crisis}. At the heart of reproducibility is a judgment about whether two results are indeed the same. However, what is used to make this judgment can differ based on the type of result and the person making the judgment. Sometimes this judgment is aided by the artifacts that are provided with the result including the data, code, and documentation. Provenance, a record of how the original result was obtained, is an important piece of information that can inform such judgments. In this paper, we argue that while provenance generally has a positive impact on the reproducibility of results, in some cases, it can complicate matters. 

Specifically, provenance can sometimes help users detect changes, but can also lack particular details or add inconsequential details, making a judgment about whether a result was reproduced may be more difficult. For example, Fig.~\ref{fig:published} shows a visualization that was published along with the code and data to reproduce the visualization. However, even with the same version of the plotting library, another user may generate a visualization with a different appearance using the same code~(Fig.~\ref{fig:reproduced}). In this case, the user had set their default configuration to a different font, colormap, and axis constraints, and the provenance did not capture the original defaults. Here, the difference is generally \emph{cosmetic} as the changes may not affect the insights obtained from the visualization. In Fig~\ref{fig:modified-data}, a casual glance at the visualization would lead one to conclude that the results are the same, despite one data item being removed (circled in red). Here, provenance \emph{should} capture the changed data source and alert a user even if the change is hard to see.  Depending on the detail of the provenance, a slight change to the process used in creating the visualization (e.g. using an alias of the original plotting command) may encourage a user to conclude that because the process was different, the results were not reproduced despite the visualizations being exactly the same. Finally, Fig.~\ref{fig:modified-process} matches Fig.~\ref{fig:published} exactly, but it may have been generated by different code or only arrived at after some user actions.
When visualizations are interactive, the initial view and its configuration may not match views explored in another session. Provenance can capture these interactions, but in general, the assessment of the reproducibility of a result is not impacted by such exploration. However, because the insights obtained may differ, perhaps this should be a consideration.

Because data is the input for visualization, any modifications to the data captured through provenance can be applied to understand how the resulting visualization changed. In many systems, data and visualizations are linked; changes to the data trigger updated visualizations~\cite{msexcel,lee:lux}. There has also been work to make sure that interactions on data spreadsheets~\cite{brachmann:vizier} and visual interfaces~\cite{psallidas:provenanceInteractiveVisualizations,wu:b2} are written to provenance and/or code. This might also function in reverse where changes in the data display meaningful cues in the visualizations. There exists work to highlight interaction histories in the visualizations themselves~\cite{feng:hindsight}, and these can be tied to provenance information as well~\cite{chakhchoukh:understandingHowInVisualization}. We believe these approaches can be enhanced, however, to better draw attention to factors of reproducibility in visualizations, data, and insights. With more fine-grained provenance of data, we might not only detect when a visualization changes but also encode this in the visualizations themselves. Similarly, when interactions on visualizations affect the perception of data items, pushing this provenance back to the data may allow us to compute exactly which data is affected.

We argue that while provenance usually aids in reproducibility, we need to be careful about how the two are connected. First, two identical visualizations need not have the same provenance. Second, two identical provenance traces can be associated with visualizations that are different. Finally, even when visualizations are identical (different), viewers can arrive at different (identical) conclusions. Therefore, we should be prudent in using provenance to evaluate reproducibility.

\section{Provenance in Computations and Visualizations}

Provenance has received considerable attention in the past couple of decades and has a vast array of uses~\cite{herschel:prov-survey}. Provenance has also been refined to be meaningful in particular domains, including in databases~\cite{buneman:dbprov,cheney:dbprov} and visualization~\cite{ragan:prov-survey,xu:prov-analysis-survey}. Computational provenance focuses on the data lineage and processes involved in a result~\cite{freire:prov-survey, simmhan:provenance} while insight provenance focuses on the rationale for decisions~\cite{gotz:insight-prov}. The specialization of provenance has followed core questions in the respective fields and allowed the development of different techniques. We consider four classes of provenance:

\begin{itemize}
\item \emph{Data provenance} concerns the lineage of data, including the inputs that contributed to a particular data item. In databases, fine-grained provenance traces individual tuples through relational operations~\cite{buneman:dbprov}.
\item \emph{Process provenance} concerns the steps involved in obtaining a result, including visualizations. This can be specified by code or a workflow and captured as an execution log~\cite{freire:prov-survey}.
\item \emph{Interaction provenance} captures the steps a user makes as they interact with a representation of data~\cite{psallidas:provenanceInteractiveVisualizations, ragan:prov-survey}. Note that this need not be limited to visualizations as state can also be captured (e.g.~\cite{camisetty:simprov}).
\item \emph{Insight provenance} traces the reasons why particular conclusions were reached, going beyond what the computer does to understand how people made a decision~\cite{gotz:insight-prov}.
\end{itemize}

\noindent Data and process provenance cover a wide domain and are often independent of visualization, while interaction and insight provenance are often more closely associated with visualization. However, all of these classes can be tied to both data and visualizations. Filtering in a visualization can be characterized as a transformation of the data and represented as data provenance~\cite{wu:b2}. Similarly, interactions with a spreadsheet may also be important details to capture to understand how data is manipulated~\cite{brachmann:vizier}.

In the database community, where data is composed of tuples, different types of provenance have been distinguished including why, how, and where~\cite{cheney:dbprov}. These relate to the tuples that contribute to a resulting tuple (why), detail about the operations involved in the result (how), and the original locations of the resulting data (where). A number of variations on this questions theme have been adopted to characterize provenance more generally~\cite{herschel:prov-survey} and in visualization~\cite{ragan:prov-survey, xu:prov-analysis-survey}. These characterizations are important because they point out the diversity of provenance information and how the differences facilitate use. Importantly, the use of provenance extends well beyond simply capturing the past or reproducing a past result. Provenance can be used to build models of user activity, suggest potential actions, and understand user intent~\cite{xu:prov-analysis-survey}.

\section{Reproducibility}

The National Academy of Science's report on reproducibility states that a result is \emph{reproducible} if one obtains ``consistent computational results using the same input data, computational steps, methods, code, and conditions of analysis''~\cite{nas:reproducibility}. Note that it focuses on the result because the computational steps are assumed to be the same. In practice, this may be difficult to guarantee due to differences in hardware or configuration. While containers and virtual machines can mitigate many of the issues, this often comes at the expense of being able to reuse published results in standard work environments. In visualization, the task of reproducing a result is complicated by the different types of results spanning a wide variety of areas~\cite{fekete:exploringReproducibilityVisualization}.

We focus on the reproducibility of visualizations as artifacts, a combination of the system that displays the visualization and the encoding itself.
From one perspective, reproducibility is the same whether the result is data displayed as a table of statistics or as a visualization. For tables, we care about whether the values are the exactly same or in the case of floating-point values, within some tolerance. For visualizations, we might similarly check whether the underlying graphical marks (e.g. pixels or vector elements) match. In many cases, such judgments can be automated. We can check that the tables do indeed match up without painstakingly comparing them manually. We can do the same for visual elements, but note that just as statistics may mask important patterns in the data, visualizations can also mask differences.

We can also distinguish between the reproducibility of a static visualization, generally an image, versus an interactive visualization, that usually lives in the context of a visualization system or framework.
For static visualizations, there is less involved in comparing visualizations because the elements are fixed. Most of the work here is focused on the pipeline that generates the final visualization. For interactive visualizations, the user can perform a number of operations that change the appearance of the visualization over time. Importantly, these actions can result in different views but \emph{different sequences} of actions can produce the same view (e.g. Fig.~\ref{fig:modified-process}). Certain actions may be reordered without any impact to the final view, and totally different actions may result in the same final view. To borrow from the language of database provenance, the why provenance may be the same, but the how provenance differs.

\section{Impacts of Provenance on Reproducibility}

We believe that understanding how provenance impacts reproducibility in visualization is important in understanding how to best capture and use provenance. In this section, we present some questions and ideas related to this interplay:
\begin{itemize}
    \item When does visualization fail, and how can provenance help?
    \item Can we use the underlying data and its provenance to characterize similarities or differences in visualizations?
    \item Is it possible to ignore certain cosmetic differences between visualizations when evaluating their similarity?
    \item How detailed must captured provenance be to ensure reproducibility? 
    \item How should we integrate interaction provenance with data and process provenance?
    \item Can we integrate provenance into visualizations?
\end{itemize}

\paragraph{Provenance Importance}
At the end of the day, the goal of visualization is not to generate particular pixels on the screen but rather to communicate information to a viewer.
A visualization can fail in this respect when the viewer comes to conclusions that are incorrect. Sometimes, this can be due to the visualization literacy of the viewer~\cite{borner:literacy}, but it can also be due to issues with the visualizations themselves. The types of problems highlighted in the algebraic treatment of visualization~\cite{kindlmann:algebraic-vis} and their characterization as visualization mirages~\cite{mcnutt:mirages} highlight a step beyond standard judgments in reproducibility.
McNutt~et~al. note that a better understanding of the input data and analytical process can help viewers address such mirages~\cite{mcnutt:mirages}, hinting at the importance of going beyond the final result and examining the provenance. Especially when a user did not generate the original visualization, understanding the process can help inform the conclusions that should or should not be made.

\paragraph{Examining Underlying Data}
In some cases, we might ask whether the reproducibility of a visualization should be characterized by the underlying data \emph{instead of} the visualization itself. Indeed, much work has focused on extracting the underlying data from visualizations in order to allow for further analyses or re-visualization (e.g.~\cite{harper:deconstructingRestylingD3,jung:chartsenseInteractiveData,luo:chartocrDataExtraction}). With the process provenance that captures the data transformation steps from the initial data to the data used for visualization, these potentially error-prone methods are unnecessary. Figure~\ref{fig:modified-data} shows that detecting a change in the visualization may be simplified by looking at the input data. But how should we deal with data that is occluded? Suppose the point that was removed was not visible. The visualization will not change at all. Even when the underlying data is the same, as in Figure~\ref{fig:reproduced}, since the viewer is seeing the visualization and not the data, it seems like only relying on the data produces an incomplete assessment.

\paragraph{Cosmetic Differences}
In some cases, \emph{cosmetic differences} like the font used or presence of absence of grid lines may be viewed as inconsequential in analyzing the reproducibility of a visualization. A pixel-by-pixel matching of outputs should probably be viewed as overkill in evaluating reproducibility. At the same time, changes like the colormap or symbol alphabets can affect perception and given the effectiveness of particular encodings, we cannot discount all such differences.

\paragraph{Provenance Granularity}
In some cases where provenance complicates an assessment of reproducibility, the issue may not be that having provenance is problematic but rather that the implementations of provenance capture are often incomplete or too coarse. Capturing every detail may allow us to detect when a visualization differs but it may not help us explain why it differs. At the same time, the overhead in storing such fine-grained detail may be prohibitive and impact the latency of the visualization. Furthermore, even if fine-grained provenance is captured, it often requires significant work to distill it to an understandable form. Grammars of interaction may help here~\cite{gathani:grammarBasedApproach}. Solutions to chunk~\cite{heer:graphicalHistoriesVisualization}, query~\cite{stitz:knowledgepearlsProvenanceBasedVisualization}, and display~\cite{walchshofer:provectoriesEmbeddingbasedAnalysis} the provenance help but in practice, many tools select a coarser provenance, knowing that every last detail cannot be captured. 

\paragraph{Relating Interaction and Process Provenance}
In the context of interactive visualizations, the ability to translate between interactions on the visualization and transformations of the data provides an important path both for capturing the provenance of interactions and comparing the reproducibility of them (i.e. if the actions are replayed~\cite{camisetty:simprov, cutler:trrackLibraryProvenanceTracking}). Work that allows actions to be translated from visualizations to code or data provenance can be helpful in capturing this type of information~\cite{psallidas:provenanceInteractiveVisualizations,wu:b2}. At the same time, certain changes or navigation operations may not have encodings in the data.

\paragraph{Making Provenance Visible}
There are also opportunities to display provenance near or in the visualizations themselves. 
These range from graphs encoding the steps followed with different glyphs~\cite{stitz:knowledgepearlsProvenanceBasedVisualization} to screenshots displaying the state of the visualizations~\cite{camisetty:simprov}.
Showing users what has been viewed affects their explorations~\cite{feng:hindsight} and in-visualization provenance views represent methods to help users better understand their work~\cite{chakhchoukh:understandingHowInVisualization}.
Such solutions complicate both the visual encoding, and potentially the reproducibility, as the encoded provenance of interaction changes over time.

\section{Conclusion}

There are a number of open questions surrounding how provenance relates to reproducibility, and the judgments surrounding the reproducibility of a visualization are both enhanced and complicated by this information. Checking if data was reproduced can be cleaner because we can ignore presentation characteristics, but because visualizations are often used as the core artifact for conclusions, we cannot totally ignore encoding characteristics. In some cases, reproducibility is aided by greater detail in provenance and stronger links from user interactions to provenance. One opportunity to better combine provenance and visualization is to visualize the provenance. In the future, we should seek to better connect captured provenance to visualization being mindful that provenance often does not guarantee reproducibility.

\bibliographystyle{abbrv-doi-hyperref}

\bibliography{paper}

\end{document}